\documentclass[conference]{IEEEtran}
\usepackage{cite}

\ifCLASSINFOpdf
\else
\fi
\usepackage{amsmath}
\usepackage{amssymb}
\usepackage{mathtools}
\usepackage{url}

\usepackage[dvipsnames]{xcolor}
\usepackage{hyperref}

\usepackage[capitalize,nameinlink]{cleveref}
\crefname{figure}{Fig.}{Figs.}
\Crefname{figure}{Figure}{Figures}
\crefname{section}{\S}{\S\S}
\Crefname{section}{\S}{\S\S}
\crefname{subsection}{\S}{\S\S}
\Crefname{subsection}{\S}{\S\S}
\crefname{subsubsection}{\S}{\S\S}
\Crefname{subsubsection}{\S}{\S\S}

\usepackage{listings}
\usepackage{tikz}
\usetikzlibrary{positioning,arrows.meta,shapes.geometric,calc,bending}
\definecolor{dkgreen}{rgb}{0,.6,0}
\definecolor{dkblue}{rgb}{0,0,.6}
\definecolor{dkyellow}{cmyk}{0,0,.8,.3}

\definecolor{keywordcolor}{rgb}{0.7, 0.1, 0.1}   
\definecolor{tacticcolor}{rgb}{0.0, 0.1, 0.6}    
\definecolor{commentcolor}{rgb}{0.4, 0.4, 0.4}   
\definecolor{symbolcolor}{rgb}{0.0, 0.1, 0.6}    
\definecolor{sortcolor}{rgb}{0.1, 0.5, 0.1}      
\definecolor{attributecolor}{rgb}{0.7, 0.1, 0.1} 

\definecolor{isarblue}{HTML}{006699}
\definecolor{MLblue}{HTML}{00334D}
\definecolor{isarlight}{HTML}{0099FF}
\definecolor{isargreen}{HTML}{009966}
\definecolor{isarpurple}{HTML}{800080}
\lstdefinelanguage{isar}{%
    keywords=[1]{type_synonym,datatype,fun,abbreviation,definition,lemma,theorem,corollary,have,by,obtain,consider,let,\{,\},apply,apply_end,done,subgoal,define,write,interpret,note},
    keywordstyle=[1]\bfseries\color{isarblue},
    keywords=[2]{where,assumes,shows,and,if,for,premises},
    keywordstyle=[2]\bfseries\color{isargreen},
    keywords=[3]{else,case,of,SOME,in,O,proof,qed,show,assume,fix,next},
    keywordstyle=[3]\bfseries\color{isarlight},
    keywords=[4]{with,then,thus,hence,from,using,also,finally,moreover,ultimately},
    keywordstyle=[4]\bfseries\color{isarblue},
showstringspaces=false,
keepspaces=true,
columns=[l]flexible,
}

\lstdefinelanguage{MiniLang}{%
    keywords=[1]{theorem,lemma},
    keywordstyle=[1]\bfseries\color{isarblue},
    keywords=[2]{WITH,WITHOUT,where,and,VARS,arbitrary,rule},
    keywordstyle=[2]\bfseries\color{isargreen},
    keywords=[3]{RULE,LET,CONSIDER,UNFOLD,CASE_SPLIT,INDUCTION,CHOOSE,APPLY,INTRO,HAVE,END,NEXT,SIMP,CONFIG,OPEN,SIMPLIFY,NOTATION,BRANCH},
    keywordstyle=[3]\bfseries\color{MLblue},
columns=[l]flexible,
}

\newcommand{\minilang}[1]{\lstinline[language=MiniLang]{#1}}

\lstdefinestyle{XXX}{
basicstyle=\ttfamily\footnotesize,
moredelim=**[is][]{@}{@},
escapeinside={&}{&}
}

\definecolor{jsonPunct}{HTML}{707070}  
\definecolor{yamlKey}{HTML}{0033B3}    
\definecolor{jsonVal}{HTML}{50A14F}    
\lstdefinelanguage{json}{
  showstringspaces=false,
  breaklines=true,
  basicstyle=\ttfamily,
  literate=
    {\{}{{\textcolor{jsonPunct}{\{}}}{1}
    {\}}{{\textcolor{jsonPunct}{\}}}}{1}
    {[}{{\textcolor{jsonPunct}{[}}}{1}
    {]}{{\textcolor{jsonPunct}{]}}}{1},
  morestring=[b]",
  stringstyle=\color{yamlKey}\bfseries,   
  comment=[l]{:},                         
  commentstyle=\color{black}\mdseries,    
  keywords={true,false,null},
  keywordstyle=\color{black},
}
\lstdefinelanguage{yaml}{
  basicstyle=\ttfamily,
  identifierstyle=\color{yamlKey}\bfseries,  
  sensitive=true,
  showstringspaces=false,
  breaklines=true,
  comment=[l]{:},                            
  commentstyle=\color{black}\mdseries,       
  literate={-\ }{{\textcolor{jsonPunct}{-}\ }}{2},
}
\usepackage{tikz}
\usetikzlibrary{positioning,decorations.markings,arrows.meta}
\usepackage[edges]{forest}
\usepackage{wrapfig}
\usepackage{booktabs}
\usepackage{multirow}

\DeclareMathOperator{\Slot}{Slot}
\DeclareMathOperator{\Opr}{Opr}
\DeclareMathOperator{\Goal}{Goal}

\newcommand{\xeval}[1]{\langle#1\rangle}
\usetikzlibrary{calc}

\definecolor[named]{ACMPurple}{cmyk}{0.55,1,0,0.15}
\hypersetup{
    colorlinks=true,
    linkcolor=ACMPurple,
    filecolor=ACMPurple,
    citecolor=ACMPurple,
    urlcolor=cyan,
    pdftitle={Agent over Abstract Syntax Tree},
    pdfpagemode=FullScreen,
    }
\DeclareMathOperator{\xlen}{len}

\usepackage{graphicx}
\newcommand{\HugeLang}[2]{%
  \mathopen{\vcenter{\hbox{\scalebox{#1}[#2]{$\langle$}}}}%
}

\newcommand{\HugeRang}[2]{%
  \mathclose{\vcenter{\hbox{\scalebox{#1}[#2]{$\rangle$}}}}%
}

\usepackage{caption}
\usepackage{subcaption} 

\newcommand{\llangle}{\langle\!\!\langle}
\newcommand{\rrangle}{\rangle\!\!\rangle}

\begin{document}
%
\title{Theorem-Proving Agent over Abstract Syntax Tree
}


\author{\IEEEauthorblockN{Qiyuan Xu}
\IEEEauthorblockA{Nanyang Technological University\\
Singapore\\
qiyuan.xu@ntu.edu.sg}
\and
\IEEEauthorblockN{Joshua Ong Jun Leang}
\IEEEauthorblockA{Imperial College London\\
London, UK\\
j.ong25@imperial.ac.uk}
\and
\IEEEauthorblockN{Renxi Wang}
\IEEEauthorblockA{MBZUAI\\
Abu Dhabi, UAE\\
renxi.wang@mbzuai.ac.ae}
\and
\IEEEauthorblockN{Wenda Li}
\IEEEauthorblockA{University of Edinburgh\\
Edinburgh, UK \\
wenda.li@ed.ac.uk}
\and[\hfill\mbox{}\par\smallskip\mbox{}\hfill]
\IEEEauthorblockN{Haonan Li}
\IEEEauthorblockA{MBZUAI\\
Abu Dhabi, UAE\\
haonan.li@mbzuai.ac.ae}
\and
\IEEEauthorblockN{Luke Ong}
\IEEEauthorblockA{Nanyang Technological University\\
Singapore\\
luke.ong@ntu.edu.sg}
\and
\IEEEauthorblockN{Conrad Watt}
\IEEEauthorblockA{Nanyang Technological University\\
Singapore\\
conrad.watt@ntu.edu.sg}
}


%


\maketitle

\begin{abstract}

Interactive theorem proving (ITP) underpins program verification and formalized mathematics, but its manual effort limits scalability. LLM-based proof agents promise to ease this effort, but their heavy token consumption and API cost remain a major obstacle. We trace this cost to a shared root: current agents operate on serialized concrete syntax, emitting proofs as source text and recovering proof states through separate, line-number-based queries, so every edit shifts later lines and forces repeated relocation of errors and states. This same dependence on concrete syntax also blocks adoption of Minilang, a recent proof language that reaches SOTA on LLM-based proving but is too new for LLMs' training corpora. We address both problems by lifting the agent off source text and onto the abstract syntax tree (AST): the model supplies proofs as JSON representations of Minilang's AST---native to tool-calling LLMs---and drives the prover through a tree-edit model that fuses proof operations and states into one proof tree, so each operation carries its own subgoal's state, readable directly off the tree. We realize this design in \emph{Agent over AST} (AoA). Against Amazon's Isabelle Agent on miniF2F and NTP4VC-Pearl common success sets, AoA cuts API cost by 2.3--4.7$\times$ (normalized input-cache accounting), uses 2.9--6.9$\times$ fewer tokens and 3.9--8.9$\times$ fewer tool calls, and finishes 1.4--2.0$\times$ faster, while also solving far more problems on the harder verification benchmark.
Across proof languages, AoA sets a new state of the art on NTP4VC-Pearl, with an 89.2\% pass rate. On miniF2F with GPT~6~sol, AoA attains 100\% pass@1 at an average of \$0.17 per case, making AoA the most cost-efficient agent to solve the benchmark in full, to our knowledge.

\end{abstract}


%
\IEEEpeerreviewmaketitle

\section{Introduction}\label{sec:Intro} 

Theorem proving is fundamental to program verification and formalized mathematics.
Interactive Theorem Proving (ITP), with its greater expressiveness compared to search-based Automated Theorem Provers (ATPs), serves as the foundation for advanced program verifiers featuring rich program logics~\cite{polu2020generative,yang2023leandojo,sammler2021refinedc,10.1145/3704903, xu_2026_22069509} and for the formalization of sophisticated mathematical theories~\cite{chen2025seedprover15,Deepseek-Prover-V2}.
However, despite this expressiveness, program verifiers and formalized mathematical projects built on top of ITPs are widely criticized for their tremendous labor cost, which limits their scalability significantly.

Promisingly, emerging Large Language Model (LLM) agents~\cite{AxAgent,Numina-Lean-Agent,MerLean,rao25a,chen2025seedprover15,Deepseek-Prover-V2} provide powerful automation for ITPs, and thereby offer an opportunity to address this limitation, transforming the advanced verification and formalization techniques confined to academia into software engineering methods that can be adopted pervasively.
For example, on the benchmark VeriSoftBench~\cite{xin2026verisoftbench} consisting of proof goals from projects about compiler correctness, programming language semantics, and verification frameworks, the leading agent Aleph prover~\cite{logicalintelligence2025aleph} generates Lean proofs achieving 94\% pass rate~\cite{logicalintelligence2025aleph}.
On PutnamBench~\cite{tsoukalas2024putnambench}, a formalization of the graduate-level Putnam mathematics competition~\cite{putnam}, Aleph prover achieves a pass rate of 99.4\%~\cite{putnambench_leaderboard}.

However, these advances also expose a general weakness among existing proof agents: the cost of API calls to LLMs is prohibitively high.
The Aleph prover incurs an average cost of \$68 per problem on PutnamBench; AxProverBase~\cite{requena2026minimal}, which achieves only a 54\% pass rate on PutnamBench, averages \$12.6 per problem across multiple benchmarks; the Numina Lean Agent~\cite{Numina-Lean-Agent} budgets roughly \$50 per problem on the Putnam 2025 competition, rising to \$1{,}000 for its hardest problems; and Archon~\cite{Archon} spends \$975 on average to prove a single research-level math problem.
%
%
It should also be noted that concrete program verification tasks may be substantially more expensive than the figures reported on the mathematical benchmarks above, for two reasons:
First, verifying a program, even just its critical components, involves a large number of proof goals. The verification of $\sim$10k lines of C code in the seL4 OS kernel involves $\sim$70k proof goals.
Second, a typical program verification process frequently revises the formalization, either to correct flaws in it or in response to other factors such as code updates. As a result, the seL4 repository records $\sim$140k distinct proof goals across its commit history.
Pricing each proof goal at AxProverBase’s reported \$12.6 average cost per problem across benchmarks, seL4’s $\sim$70k–140k proof goals imply roughly \$0.9M–\$1.8M in API cost alone.
Note that seL4 is an academic project; for industrial-scale applications, the cost could be far greater.
To cut costs, one could certainly switch to cheaper models. However, given the tension between cost and the reasoning capability of language models, one would naturally prefer to use powerful models at a lower cost. This raises our research question:
\begin{list}{}{\leftmargin=1em \rightmargin=1em \topsep=0.5\baselineskip \parsep=0pt \itemsep=0pt}
\item[\textbf{RQ1.}] How can we reduce the API cost of proof agents, from the side of agent design?
\end{list}

In parallel, a recent work introduced Minilang, a redesigned proof language aimed at improving LLM efficiency on theorem proving. Minilang~\cite{Minilang} emphasizes that LLMs should focus on high-level proof design while delegating the fine-grained logical reasoning they are poor at to classical ATPs with well-defined search-based algorithms.
In addition, Minilang redesigns its language constructs to provide LLMs with more automated and higher-level interfaces that mitigate the gap between formal proofs and the natural-language corpora that general-purpose LLMs are trained on.
We believe that these designs can reduce the reasoning burden on LLMs and improve the accuracy of their proof generation, thereby effectively lowering the API cost of proof agents.

However, one difficulty is that Minilang is a new language whose large proof corpus was released only on Dec. 22, 2025.
Agents cannot rely on general-purpose LLMs having corpus-scale familiarity with Minilang proofs or its concrete syntax.
Moreover, fine-tuning on such a corpus may be theoretically feasible, but it is expensive in practice and faces obstacles such as data scarcity for newly designed languages and catastrophic forgetting. This leads to our second research question:
\begin{list}{}{\leftmargin=1em \rightmargin=1em \topsep=0.5\baselineskip \parsep=0pt \itemsep=0pt}
\item[\textbf{RQ2.}]
Can we enable general-purpose LLMs to construct formal proofs in a language for which they lack corpus-scale proof exposure, purely through system design and without modifying the underlying models?
\end{list}

Our answer rests on a common practice in programming language studies: concrete syntax is largely incidental, so one abstracts away from it and defines a language's syntax as an Abstract Syntax Tree (AST).
In this paper, we present and study a proof agent built on Minilang's AST.
In detail, we define Minilang's AST as a recursive algebraic datatype and formalize it using JSON Schema, a format that LLMs are natively and extensively trained on to support tool calling.
With this JSON Schema, LLMs can then express Minilang's ASTs as JSON representations.
This removes the LLMs' dependence on any particular concrete syntax of Minilang, allowing them to focus on how a proof should be constructed semantically, rather than on how it must be spelled out in that syntax.
Moreover, general-purpose LLMs are capable of this semantic construction because Minilang is deliberately designed so that each of its language constructs mirrors a step in natural-language, pen-and-paper reasoning.
Essentially, we abstract away from the concrete syntax, extract the semantic information it conveys, and re-express it in LLMs' native languages.
In this way, we provide a concrete solution to RQ2 for the case of Minilang.

The above addresses how to get LLMs to produce Minilang. A proof agent, however, does more than have the LLM emit proof scripts; it also requires intricate interaction with the ITP system, such as retrieving proof states and error messages.

In conventional proof agents~\cite{AxAgent,Archon,Numina-Lean-Agent,MerLean,CoqPilot,AutoCorrode}, LLMs generate proofs by writing source-code snippets into a plain-text file through a tool call.
After completing a proof edit, the agents typically require the LLMs to issue another tool call to run the proof checker.
If the proof fails to check, the agents' responses to the LLMs locate each error in the proof by a file path and a line number.
The LLMs must then navigate to the reported location and read the surrounding source code with a read tool. They may also need additional Language Server Protocol (LSP) tool calls, for example, to inspect the proof state or the type of a variable at a given location, in order to understand why the proof fails and determine how to repair it.
Consequently, the process incurs a large number of tool calls. This is a major contributor to the high API cost, because each tool call constitutes a separate API request that must carry the entire conversation history as input.

We observe that this inefficiency shares the same root cause as Minilang's generation problem above: the agent is forced to operate on serialized concrete syntax. Because the proof lives as plain text, the proof state, errors, and types the LLM needs are not directly available, and must be recovered through separate, line-number-based tool calls. We therefore apply the same remedy to the interaction as a whole.
Just as in our earlier ``abstract away and re-express'' approach, we extract the abstract semantic model underlying the interactive proving process and formulate it as a novel \emph{tree-edit proof model}, which we believe better recovers a proof's natural deductive structure, in the spirit of Natural Deduction, long regarded as close to the way humans naturally construct proofs.
In this model, the evolving proof state and the proof operations are fused into a \emph{proof tree}, so that every proof operation carries the state of its own subgoal. 
The agent advances the proof by editing the tree directly rather than by patching source code text. This gives a single, self-contained representation that removes the indirection of conventional agents: everything the LLM needs is read directly off the tree, rather than chased through brittle line numbers that shift with every edit.


%


Based on this design, we present \emph{AoA}, a proof agent built on Minilang.
AoA substantially reduces the cost of proof agents.
On the common success sets, and under normalized input-cache-rate accounting, AoA achieves 2.3--4.7$\times$ lower API cost than Amazon’s Isabelle Agent. It also reduces total token consumption by 2.9--6.9$\times$. This saving is accompanied by a 3.9--8.9$\times$ reduction in the number of tool calls, and AoA is not slower: in our runs, it finishes the cases 1.4--2.0$\times$ faster.
These results support our answer to RQ1: redesigning the proof language and exposing its AST within a tree-edit proof model effectively reduces proof-agent cost.


Beyond these efficiency gains, AoA also achieves strong absolute proving performance: it reaches an 89.2\% pass rate on the NTP4VC-Pearl benchmark, setting a new state of the art for verification-condition proving, and achieves 100\% on miniF2F.
We summarize the contributions of this paper.

\begin{itemize}
    \item We propose a tree-edit proof model over Minilang AST, replacing source-file patching with proof-tree editing and eliminating the need for separate line-number-based proof-state queries.

\item We present AoA, a proof agent built on this tree-edit model, enabling general-purpose LLMs to construct proofs in a proof language without corpus-scale exposure to its proof corpus, and without fine-tuning the LLM.

    \item We evaluate and analyze AoA on miniF2F and NTP4VC-Pearl against Amazon's Isabelle Agent.
\end{itemize}

\section{Background}

This section first provides the necessary background on Interactive Theorem Proving (ITP --- \cref{sec:itp}). It then reviews the mainstream design of current-generation proof agents (\cref{sec:existing_agent}), which serves as a baseline against which we present our novelty. Finally, it introduces Minilang (\cref{sec:minilang}), the proof language on top of which our agent AoA is built.

\subsection{Interactive Theorem Proving}\label{sec:itp}

ITP develops a proof as a dialogue between the user and the prover. The user supplies the proof one step at a time, and the prover responds with the resulting \emph{proof state}: the goals that remain to be proved, together with the hypotheses available for proving them. Reading this state, the user decides on the next step, and the proof grows through this back-and-forth.

A subtlety is that the proof state at a given point is generally not evident from the proof script itself; it has to be computed by the prover. To inspect it, users typically move the cursor to the position of interest, and the prover reports the proof state holding at that point. This cursor-driven, position-based access to the proof state is the interaction model that current proof agents inherit.

\subsection{The Conventional Paradigm of Proof Agents}\label{sec:existing_agent}

A proof agent equips a language model with tools for editing and checking proofs. Current-generation theorem-proving agents largely follow the coding-agent paradigm: the model edits proof source code until it is accepted by the checker. This source-centered design has two limiting features:
(1) the language model directly produces the proof in the concrete syntax of the target proof language; and (2) most interactions with the prover are built on source code positions (file path, line, and column).

Two representative examples are the Numina Lean Agent~\cite{Numina-Lean-Agent} and Amazon's Isabelle Agent~\cite{AutoCorrode}.
The former drives Claude Code to edit Lean source files directly; the latter lets the language model patch Isabelle proof text into the source file.
To let the language model interact with the prover, both expose a series of tools, many of which are based on source-code positions. For example, to access the proof state at a given position, the Numina Lean Agent prompts the language model to specify exact file paths, line numbers, and column numbers (note that language models are known to be poor at counting the letters in a word and therefore column numbers~\cite{fu2024whycount, edman2024cute});
Amazon's agent retrieves proof states through a cursor-based procedure, requiring separate tool calls to move the cursor and inspect the state.
As tool-call responses sent back to the language model, the proof checker's error messages are passed essentially verbatim;
the model must then use the line numbers they contain to locate each error in the source file, read the surrounding code, and often issue a further tool call to inspect the proof state and context.

We see two limitations with this paradigm, one with each of the two features above.
First, generation tied to a specific concrete syntax may be acceptable for languages that general-purpose LLMs have already mastered, but it is unfavorable for newly designed formal languages that continually arise in the development and study of formal languages: for such a language, either there is too little corpus to fine-tune models, or the cost is prohibitive.
This clearly hinders the progress of formal languages, and in particular our adoption of Minilang~\cite{Minilang}.
Second, interaction built on source-code positions is inefficient. To line up the prover's responses with the corresponding parts of the proof script, the agent must constantly navigate back and forth within the source files; and because editing a source file shifts the line numbers below the edit, the positions it has already located are repeatedly invalidated, further aggravating this navigation.

\subsection{Minilang, A Proof Language Specially Designed for LLM}\label{sec:minilang}

Isabelle/Minilang~\cite{Minilang} is a high-level proof-planning language tailored to large language models.
Its design is motivated by two observations about LLMs: first, LLMs are weaker at fine-grained logical reasoning than at planning a proof as a whole, and even when they can carry out such reasoning, they do so far less efficiently than procedures built on well-defined, search-based algorithms such as resolution~\cite{Resolution}, superposition~\cite{Superposition}, and DPLL(T)~\cite{DPLLT}. Second, LLMs reason far more readily in natural language than in formal languages~\cite{wu2022autoformalization, jiang2023dsp}, a disparity commonly attributed to the predominance of natural language over formal text in their training corpora. For the first, Minilang offers a family of \emph{structural commands} for laying out high-level proof plans, and delegates the proof obligation each step incurs to classical Automated Theorem Provers (ATPs), such as Satisfiability Modulo Theories (SMT) solvers (e.g., Z3~\cite{Z3} and cvc5~\cite{CVC5}) and first-order provers (e.g., E~\cite{Eprover} and Metis~\cite{Metis}). For the second, Minilang provides a second kind of command that mirrors the operations commonly found in natural-language, pen-and-paper proofs, such as unfolding and simplification.
%


The discussion in this paper need not cover every Minilang command; we therefore leave the details to the original work~\cite{Minilang} and, following the two kinds above, formalize:
\begin{align*}
  \mathrm{Minilang\_Proof} \Coloneqq&~ \mathrm{Minilang\_Cmd}^+ \\
  \mathrm{Minilang\_Cmd} \Coloneqq&~ 
  \mathit{Op}(\mathit{args})\ \mathrm{Proof\_Blocks}\ \mbox{\minilang{END}}\\
  \mathrel{|}&~ 
  \mathit{Op}(\mathit{args})\\
  \mathrm{Proof\_Blocks} \Coloneqq&~ \mathrm{Minilang\_Cmd}^*\\*
  \mathrel{|}&~\mathrm{Minilang\_Cmd}^*\ \mbox{\minilang{NEXT}}\ \mathrm{Proof\_Blocks}
\end{align*}
A Minilang proof script is a non-empty sequence of Minilang commands (denoted as $\mathrm{Minilang\_Cmd}^+$).
A structural command (e.g., line 3 and 8 in \cref{fig:minilang_example}) is commenced by its operator $\mathit{Op}$ together with its arguments $\mathit{args}$, and terminated by the keyword END that marks the end of the proof blocks it opens.
A proof block can contain zero or more commands ($\mathrm{Minilang\_Cmd}^*$).

\begin{table}[t]
\renewcommand{\arraystretch}{1.2}
    \caption{Core tools of AoA agent}
    \label{tab:tools}
    \centering
    \begin{tabular}{l p{6.5cm}}\toprule
       \bf Name  & \bf Description \\\midrule
    \textsc{Read} & Returns the proof tree in YAML format. Optionally takes a node identifier and returns the subtree rooted at it.\\
    \textsc{Edit} & Fills an empty slot with Minilang ASTs, inserts ASTs before a node, amends an existing node, or deletes one.\\
\textsc{Query} &
Uses natural-language descriptions to semantically search for relevant lemmas, inference rules, constants, and types, optionally filtered by structure patterns.
\\
\textsc{Subagent} &
Launches a subagent to prove a specified goal under the context scoped to that goal’s subtree.
\\
\multicolumn{2}{p{8.5cm}}{
\emph{Not listed: control tools (e.g., give up) and tools for answering the agent's clarification questions (e.g., the type of a variable), both common to agents.}
}
\\\bottomrule
    \end{tabular}
\end{table}

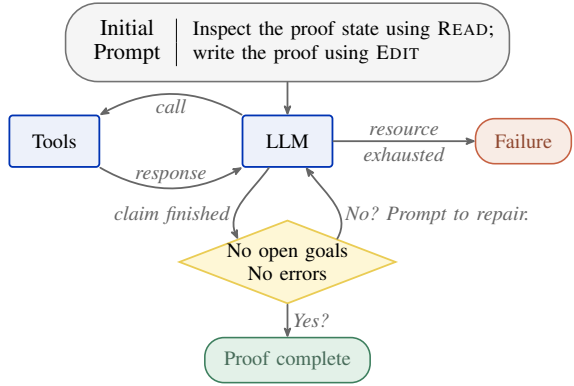
\begin{figure}
    \centering
\scalebox{0.85}{%
\begin{tikzpicture}[
    >={Stealth[round,length=2mm]},
    line cap=round,
    font=\small,
    proc/.style={rectangle, rounded corners=1.5pt, draw=yamlKey, line width=0.8pt,
                 fill=yamlKey!7, minimum width=14mm, minimum height=8mm, align=center},
    dec/.style={diamond, draw=Goldenrod!90!black, line width=0.8pt,
                fill=Goldenrod!22, aspect=2.6, inner sep=0pt, align=center},
    io/.style={rectangle, rounded corners=3mm, draw=black!55, line width=0.7pt,
               fill=black!4, minimum height=7mm, inner xsep=3mm, align=center},
    bad/.style={io, draw=BrickRed!75, fill=BrickRed!7, text=BrickRed!70!black},
    good/.style={io, draw=ForestGreen!65!black, fill=ForestGreen!10,
                 text=ForestGreen!45!black},
    arr/.style={->, line width=0.8pt, draw=black!60},
    lbl/.style={font=\small\itshape, text=black!60, inner sep=2pt, align=center},
]

\node[proc] (llm) {LLM};
\node[proc, left=22mm of llm] (tools) {Tools};
\node[io, above=5mm of llm, rounded corners=5mm, inner sep=2.4mm] (init)
      {\setlength{\tabcolsep}{2mm}\renewcommand{\arraystretch}{1.05}%
       \begin{tabular}{c@{\hskip 2.5mm}|@{\hskip 2.5mm}l@{}}
         \normalsize Initial &
           \small Inspect the proof state using \textsc{Read};\\
         \normalsize Prompt &
           \small write the proof using \textsc{Edit}
       \end{tabular}};
\node[bad, right=22mm of llm] (fail) {Failure};
\node[dec, below=8mm of llm] (check) {No open goals\\No errors};
\node[good, below=5mm of check] (done) {Proof complete};

\draw[arr] (init) -- (llm);

\draw[arr] (llm.150) to[out=150,in=30] node[below, lbl] {call} (tools.30);
\draw[arr] (tools.-30) to[out=-30,in=210] node[above, lbl] {response} (llm.210);

\draw[arr] (llm.east) -- node[above, lbl] {resource} node[below, lbl] {exhausted} (fail);

\draw[arr] (llm.235) to[out=-125,in=120]
      node[left, lbl, pos=0.68] {claim finished} (check.155);

\draw[arr] (check.25) to[out=60,in=-55]
      node[right, lbl, pos=0.32] {No? Prompt to repair.} (llm.305);

\draw[arr] (check.south) -- node[right, lbl] {Yes?} (done);

\end{tikzpicture}}
    \caption{The agent pipeline of AoA}
    \label{fig:pipeline}
\end{figure}

\newcommand{\xcomment}[1]{{\color{gray}\it#1}}

\begin{figure*}[t]
\centering
\begin{subfigure}[b]{0.49\textwidth}
\centering
\begin{lstlisting}[language=MiniLang,
style=XXX,
basicstyle=\ttfamily\fontsize{8}{9.5}\selectfont,
moredelim={**[is][{\btHL[fill=cyan!15]}]{@}{@}},
numbers=left,
numberstyle=\footnotesize\color{gray},
numbersep=5pt,
xleftmargin=0.8em
]
theorem &$\mathrm{BST\_insertion\_preserves\_elements}$&:
  &$\mathrm{elems}(\mathrm{insert}(x,t)) = \{x\} \cup \mathrm{elems}(t)$&
INDUCTION &$t$&
  &\xcomment{the base case when t is a leaf is trivial}&
NEXT
  &\xcomment{the inductive case when t is a node $\llangle l,v,r\rrangle$ with}&
  &\xcomment{left subtree $l$, value $v$, and right subtree $r$}&
  BRANCH &$x < v$& | &$x = v$& | &$x > v$&
    &\xcomment{To show the case of $x < v$, we only need a lemma:}&
    HAVE &$\mathrm{insert}(x,\llangle l, v, r\rrangle) = \llangle\mathrm{insert}(x,l), a, r\rrangle$& END
  NEXT &\xcomment{the case of $x = v$ ...}&
  NEXT &\xcomment{the case of $x > v$ ...}&
END
\end{lstlisting}
\caption{An example of Minilang}
\label{fig:minilang_example}
\vspace{1em}
\begin{lstlisting}[language=yaml,
basicstyle=\ttfamily\fontsize{7}{8}\selectfont,
escapeinside={&}{&}
]
variables: &\xcomment{\it the same as \cref{fig:initial-proof-tree}}&
goal: &\xcomment{\it the same as \cref{fig:initial-proof-tree}}&
proof:
  step id: 1
  operation: Induction
  target: &$t$&
  induction cases:
    - goal id: 1.1
      goal: &$\mathrm{elems}(\mathrm{insert}(x, \mathrm{Leaf})) = \{x\} \cup \mathrm{elems}(\mathrm{Leaf})$& 
      proof: &\text{\normalfont Empty! Call the edit tool to fill step 1.1.1}&
    - goal id: 1.2
      variables:
        v: &$\mathrm{int}$&
        l, r: &$\mathrm{int\;bst}$&
      premises:
        hyp1: &$\mathrm{elem}(\mathrm{insert}(x,l)) = \{x\}\cup \mathrm{elem}(l)$&
        hyp2: &$\mathrm{elem}(\mathrm{insert}(x,r)) = \{x\}\cup \mathrm{elem}(r)$&
      conclusion:
        &$\mathrm{elems}(\mathrm{insert}(x,\llangle l,v,r \rrangle)) = \{x\}\cup \mathrm{elems}(\llangle l,v,r \rrangle)$&
      proof: &\text{\normalfont Empty! Call the edit tool to fill step 1.2.1}&
\end{lstlisting}
\caption{The proof tree after applying line 3}
\label{fig:proof-tree-after-induction}
\end{subfigure}%
\begin{subfigure}[b]{0.49\textwidth}
\centering
\begin{lstlisting}[language=yaml,
basicstyle=\ttfamily\fontsize{7}{8}\selectfont,
escapeinside={&}{&}
]
variables:
  x: &$\mathrm{int}$&
  t: &$\mathrm{int}\;\mathrm{bst}$&
goal: &$\mathrm{elems}(\mathrm{insert}(x,t)) = \{x\} \cup \mathrm{elems}(t)$&
proof: &\text{\normalfont Empty! Call the edit tool to fill step 1}&
\end{lstlisting}
\caption{Initial proof tree}
\label{fig:initial-proof-tree}
\vspace{.5em}
\begin{lstlisting}[language=json,
basicstyle=\ttfamily\fontsize{7}{8}\selectfont,
escapeinside={&}{&}
]
{ "operation": "Induction",
  "target": "&$t$&",
  "proofs": "Given_Later"
}
\end{lstlisting}
\caption{Minilang AST at line 3, in JSON}
\label{fig:json_AST_line_3}
\vspace{.5em}
\begin{lstlisting}[language=yaml,
basicstyle=\ttfamily\fontsize{7}{8}\selectfont,
escapeinside={&}{&}
]
- step id: 1
  - goal id: 1.1
    statement: &$\mathrm{elems}(\mathrm{insert}(x, \mathrm{Leaf})) = \{x\} \cup \mathrm{elems}(\mathrm{Leaf})$& 
    msg: Unfinished proof! Fill step 1.1.1
  - goal id: 1.2
    variables: &\xcomment{variables $v,l,r$ and their types}&
    premises: &\xcomment{hypothese hyp1 and hyp2}&
    conclusion: &$\mathrm{elems}(\mathrm{insert}(x,\llangle l,v,r \rrangle)) = \{x\}\cup \mathrm{elems}(\llangle l,v,r \rrangle)$&
    msg: Unfinished proof! Fill step 1.2.1
\end{lstlisting}
\caption{The return of the Edit tool for filling line 3}
\label{fig:compressed_proof_tree}
\vspace{.5em}
\begin{lstlisting}[language=json,
basicstyle=\ttfamily\fontsize{7}{8}\selectfont,
escapeinside={&}{&}
]
{ "operation": "Branch",
  "cases": [
    { "statement": "x < v"
      "proofs": [{
        "operation": "Have",
        "statement": "&$\mathrm{insert}(\cdots) = \cdots$&"
        "proof":[{"operation": "Obvious"}]
      }]
    }, &\xcomment{other cases...}& ]
}
\end{lstlisting}
\caption{Minilang AST for line 8-14, in JSON}
\label{fig:json_AST_branch}
\end{subfigure}
\caption{Examples of Minilang, the JSON representation of its AST, and the YAML representation of the proof tree.}
\label{fig:examples}
\end{figure*}
\section{Method: Over the Abstract Syntax Tree}\label{sec:method}

As motivated in \cref{sec:Intro}, AoA interacts with the prover through a tree-edit proof model over Minilang's AST. This section develops this design. \cref{sec:running_example} first presents a running example to illustrate how AoA constructs a proof by editing the proof tree. \cref{sec:tree-edit-model} then formalizes the underlying tree-edit model, in which proof operations and prover states are interleaved in a single tree. Finally, \cref{sec:agent} describes the proof agent built on top of this model.


\subsection{A Running Example of AoA}\label{sec:running_example}

In this example, we consider a simple functional-correctness property of Binary-Search-Tree (BST) insertion: inserting an element into a BST does not lose or create any other element; it only adds the inserted element to the set of elements already stored in the tree.
\cref{fig:minilang_example} shows how this property is formalized in Isabelle/HOL and proved in Minilang. Here, $\mathrm{insert}(x,t)$ denotes the BST obtained by inserting $x$ into $t$, and $\mathrm{elems}(t)$ denotes the set of elements stored in $t$. 

Architecturally, AoA consists of a fixed agent pipeline and a collection of LLM tools; the main ones are summarized in \cref{tab:tools}.
Among these tools, \textsc{Read} and \textsc{Edit} form the core interface to the tree-edit model, and they are the two tools used throughout this example.
The pipeline is simple: once a target proof goal is given and the agent is launched, it first prompts the LLM to inspect the current proof goal using \textsc{Read}.

The \textsc{Read} tool serializes the current proof tree into a YAML representation and returns it to the LLM.
On this first invocation, the proof tree is the initial one illustrated in \cref{fig:initial-proof-tree}, which contains the contextual variables and the goal statement.

After the LLM inspects the initial tree, the pipeline prompts it to construct the proof by calling \textsc{Edit}. The \textsc{Edit} tool takes three arguments: the identifier of the node to be edited, the edit mode, and a JSON representation of a Minilang AST.
The edit modes will be elaborated in \cref{sec:tree-edit-model}; for now, it suffices to note that the initial proof tree (\cref{fig:initial-proof-tree}) indicates that the LLM should call \textsc{Edit} to fill node $1$.
Suppose the LLM constructs the Minilang proof shown in \cref{fig:minilang_example}. The first step of this construction is the Induction command at line 3, so the LLM calls \textsc{Edit}(1, \texttt{fill}, \cref{fig:json_AST_line_3}), where \cref{fig:json_AST_line_3} denotes the JSON representation of this command's AST.

After applying this \textsc{Edit} call, the proof tree becomes the one shown in \cref{fig:proof-tree-after-induction}. Importantly, the tree records not only the proof operation just inserted, but also the effect of executing that operation on the proof state. In this example, the induction on $t$ decomposes the original goal into two subgoals: the base case where $t$ is a leaf, and the inductive case where $t$ is a node. The statements of these subgoals, together with their contextual variables and premises, are displayed directly in the proof tree. The LLM therefore need not infer or guess how the proof operation changes the proof state; it can read the result computed by the ITP system.

The LLM could obtain this updated tree (\cref{fig:proof-tree-after-induction}) by calling \textsc{Read} again. More conveniently and efficiently, however, \textsc{Edit} directly returns to the LLM a compressed overview of the updated proof tree, as shown in \cref{fig:compressed_proof_tree}.
This response incrementally synchronizes the proof-state changes across the proof tree with the LLM. Proof states that remain unchanged are omitted, while successfully completed subtrees are collapsed; the full compression policy is described in \cref{sec:agent}. In this way, the LLM is informed of all relevant changes caused by the edit without having to read the entire updated tree again.

With this updated information, the LLM continues the proof by freely invoking the tools in \cref{tab:tools} as needed. For example, it may follow the \texttt{msg} fields in the compressed response shown in \cref{fig:compressed_proof_tree} and use \textsc{Edit} to fill the empty proof slot $1.2.1$ with the Minilang AST shown in \cref{fig:json_AST_branch}, and the slot $1.1.1$ with \verb|{"operation": "Obvious"}| which delegates the proof goal to ATPs.
AoA's pipeline itself does not prescribe a fixed sequence of tool calls. It intervenes only when the LLM reports that the proof is complete: it then checks whether all subgoals have been proved and whether any evaluation errors remain. If some unfinished subgoal or failed node remains, the pipeline prompts the LLM to work on the corresponding node, and this process repeats until the proof is successfully completed or the resource limits are reached.
The full AoA pipeline is illustrated in \cref{fig:pipeline}.




\subsection{A Tree-Edit Model That Interleaves States and Operations}\label{sec:tree-edit-model}

The running example above illustrates the interaction model of AoA in concrete terms. We now formalize the underlying tree-edit proof model. The model has three components: a proof tree, whose nodes represent proof operations, goals, and empty proof slots (\cref{sec:proof-tree}); a set of edit commands that modify this tree (\cref{sec:edit-model}); and an evaluation mechanism that re-computes the affected proof states after each edit (\cref{sec:proof-evaluation}).


\subsubsection{The Tree Model}\label{sec:proof-tree}
In our model, the editable object is a \emph{proof tree}, rather than a serialized proof script. 
A proof tree is rooted at a goal node. Each goal node contains a proof component, consisting of a sequence of proof-operation nodes and, if the proof is unfinished, a trailing empty slot (denoted $\mathrm{Opr}^* \Slot^?$). Conversely, each proof-operation node contains zero or more goal nodes (denoted $\mathrm{Goal}^*$) generated by applying that operation.
This gives a tree in which proof goals and proof operations are interleaved layer by layer.
\begin{align*}
  \mathrm{Proof\_Tree}\Coloneqq\hspace*{-3em}&\hspace{3em}~\mathrm{Goal}\\
  \mathrm{Node} \Coloneqq&~ \Opr \mathrel{|} \Goal \mathrel{|} \Slot\\
  \mathrm{Opr} \Coloneqq&~(\mathit{id}, \mathit{op}, \mathit{args}, \mathit{msg}, \mathit{goals}),\qquad\mathit{goals}\in \mathrm{Goal}^* \\
  \mathrm{Goal} \Coloneqq&~(\mathit{id}, \mathit{statement}, \mathit{proof}),\quad\mathit{proof} \in \mathrm{Opr}^* \Slot^?\\
  \Slot \Coloneqq&~ (\mathit{id})\hspace*{10.6em}\mathrm{len}(\mathit{proof}) > 0\\
\mathit{id} \in \mathrm{Id} \Coloneqq&~ \mathit{ind} \mathrel{|} \mathit{id}\boldsymbol{.}\mathit{ind}\qquad \mathit{ind} \in \mathrm{Index}
\end{align*}
%
Since applying a proof operation transitions the proof state, an operation node records the resulting state in its \emph{msg} field, together with any notice or warning reported during execution.
This attaches prover feedback directly to the operation node.



The design of the node identifier aims to be informative to the LLM rather than being a meaningless, opaque token.
We therefore design a node identifier $\mathit{id}$ as a dot-separated list of indexes, forming a path descending from the root to the target node, each index identifying a child one level down, thereby encoding its structural position within the tree.
For example, identifier \texttt{2.1.4} locates the 4th operation of the 1st subgoal of the 2nd operation of the top goal.

%

\newcommand{\squeezeleft}[1]{\mathmakebox[20ex][r]{#1}}

\begin{figure*}[tp]
    \centering

\begin{tikzpicture}[scale=0.9, transform shape]
  \node (FILL-LEFT) at (-8mm,0) {%
    \begin{forest} for tree={align=center, l sep=0, s sep=0.3cm}
      [$\mathit{Goal}$
        [$\cdots$]
        [$\Slot(i)$]
      ]
    \end{forest}
  };
  
  \node (FILL-RIGHT) [right=24mm of FILL-LEFT] {%
    \begin{forest} for tree={align=center, l sep=0, s sep=0.4cm}
      [$\mathit{Goal}$
        [$\cdots$]
        [$\xeval{\mathit{segs}}_i$]
      ]
    \end{forest}
  };

  \draw[->, thick] ($(FILL-LEFT.east)+(-1mm,0)$) -- ($(FILL-RIGHT.west)+(0mm,0)$)
    node[midway, above] {Fill $(i,\mathit{segs})$}
    node[midway, below=2mm, align=center] {if $\mathit{Goal}$\\is proven};

  \node (FILL-LEFT) [right=18mm of FILL-RIGHT] {%
    \begin{forest} for tree={align=center, l sep=0, s sep=0.3cm}
      [$\mathit{Goal}$
        [$\cdots$]
        [$\Slot(i)$]
      ]
    \end{forest}
  };
  
  \node (FILL-RIGHT) [right=20mm of FILL-LEFT] {%
    \begin{forest} for tree={align=center, l sep=0, s sep=0.2cm}
      [$\mathit{Goal}$
        [$\cdots$]
        [$\xeval{\mathit{segs}}_i$]
        [$\Slot(i + \xlen(\mathit{segs}))$]
      ]
    \end{forest}
  };

  \draw[->, thick] ($(FILL-LEFT.east)+(-2mm,0)$) -- ($(FILL-RIGHT.west)+(0mm,0)$)
    node[midway, above] {Fill $(i,\mathit{segs})$}
    node[midway, below=2mm, align=center] {\small if $\mathit{Goal}$\\is not proven};
  
  \node (INSERT-LEFT) at (0,-2) {%
    \begin{forest} for tree={align=center, l sep=0, s sep=0.3cm}
      [$\mathit{Goal}$
        [$\cdots$]
        [{$Opr(i,\cdots)$}]
        [$\cdots$]
      ]
    \end{forest}
  };
  
  \node (INSERT-RIGHT) [right=12mm of INSERT-LEFT] {%
    \begin{forest} for tree={align=center, l sep=0, s sep=0.2cm}
      [$\mathit{Goal}$
        [$\cdots$]
        [$\xeval{\mathit{segs}}_{i^-}$]
        [{$\xeval{Opr(i,\cdots)}$}]
        [$\xeval{\cdots}$]
      ]
    \end{forest}
  };
  
  \draw[->, thick] ($(INSERT-LEFT.east)+(-1mm,0)$) -- ($(INSERT-RIGHT.west)+(5mm,0)$) node[midway, above] {Insert $(i,\mathit{segs})$};

  \node (DELETE-LEFT) [right=9mm of INSERT-RIGHT] {%
    \begin{forest} for tree={align=center, l sep=0, s sep=0.3cm}
      [$\mathit{Goal}$
        [$\cdots$]
        [{$Opr(i,\cdots)$}]
        [$\cdots$]
      ]
    \end{forest}
  };
  
  \node (DELETE-RIGHT) [right=15mm of DELETE-LEFT] {%
    \begin{forest} for tree={align=center, l sep=0, s sep=0.2cm}
      [$\mathit{Goal}$
        [$\cdots$]
        [$\xeval{\cdots}$]
      ]
    \end{forest}
  };
  
  \draw[->, thick] ($(DELETE-LEFT.east)+(-8mm,0)$) -- ($(DELETE-RIGHT.west)+(2mm,0)$) node[midway, above] {Delete$(i,\mathit{segs})$};

  \node (AMEND-LEFT) at (0,-4.9) {%
    \begin{forest} for tree={align=center, l sep=0, s sep=0.3cm}
      [$\mathit{Goal}_0$
        [$\cdots$]
        [{$Opr(i,\cdots)$}
            [$Goal_1$ [$\mathit{Oprs}_1$]]
            [$\cdots$ [$\cdots$]]
            [$Goal_n$ [$\mathit{Oprs}_n$]]
        ]
        [$\cdots$]
      ]
    \end{forest}
  };
  
  \node (AMEND-RIGHT) [right=32mm of AMEND-LEFT] {%
    \begin{forest} for tree={align=center, l sep=0, s sep=0.2cm}
      [$\mathit{Goal}_0$
        [$\cdots$]
        [{$\xeval{\mathit{seg}}_i$}
            [$\mathit{Goal}^{\boldsymbol{\prime}}_1$ [$\xeval{\mathit{Oprs}_1}$]]
            [$\cdots$ [$\xeval{\cdots}$]]
            [$\mathit{Goal}^{\boldsymbol{\prime}}_n$ [$\xeval{\mathit{Oprs}_n}$]]
        ]
        [$\cdots$]
      ]
    \end{forest}
  };
  
  
  \draw[->, thick] ($(AMEND-LEFT.east)+(-1mm,0)$) -- ($(AMEND-RIGHT.west)+(1mm,0)$) node[midway, above] {Amend $(i,\mathit{\mathit{seg}})$}
  node[midway, below] {
    \parbox{3.2cm}{\small
    if $\xeval{\mathit{seg}}_i$ is an operation\\and
    $\xeval{\mathit{seg}}_i$ opens $n$ goals
  }} ;
  
  \node (AMEND-LEFT) [right=6mm of AMEND-RIGHT] {%
    \begin{forest} for tree={align=center, l sep=0, s sep=0}
      [$\mathit{Goal}$
        [$\cdots$]
        [{$Opr(i,\cdots)$}]
        [$\cdots$]
      ]
                                                                            \end{forest}
  };
  
  \node (AMEND-RIGHT) [right=9mm of AMEND-LEFT] {%
    \begin{forest} for tree={align=center, l sep=0, s sep=0}
      [$\mathit{Goal}$
        [$\cdots$]
        [$\xeval{\mathit{segs}}_{i^-}$]
        [$\xeval{\cdots}$]
      ]
    \end{forest}
  };
  
  \draw[->, thick] ($(AMEND-LEFT.east)+(-5mm,0)$) -- ($(AMEND-RIGHT.west)+(5mm,0)$) node[midway, above] {Amend $(i,\mathit{segs})$}
  node[midway, below] {otherwise}
  ;

%

\end{tikzpicture}
    \caption{The behaviors of the tree edit commands. Here, we use $i$ to denote an identifier instead of $\mathit{id}$.
    Notation $\xeval{\mathit{segs}}_i$ denotes the elaboration of Minilang segments $\mathit{segs}$ into nodes. Notation $i^-$ denotes an index interpolated between $i-1$ and $i$.
    }
    \label{fig:tree-edit}
\end{figure*}
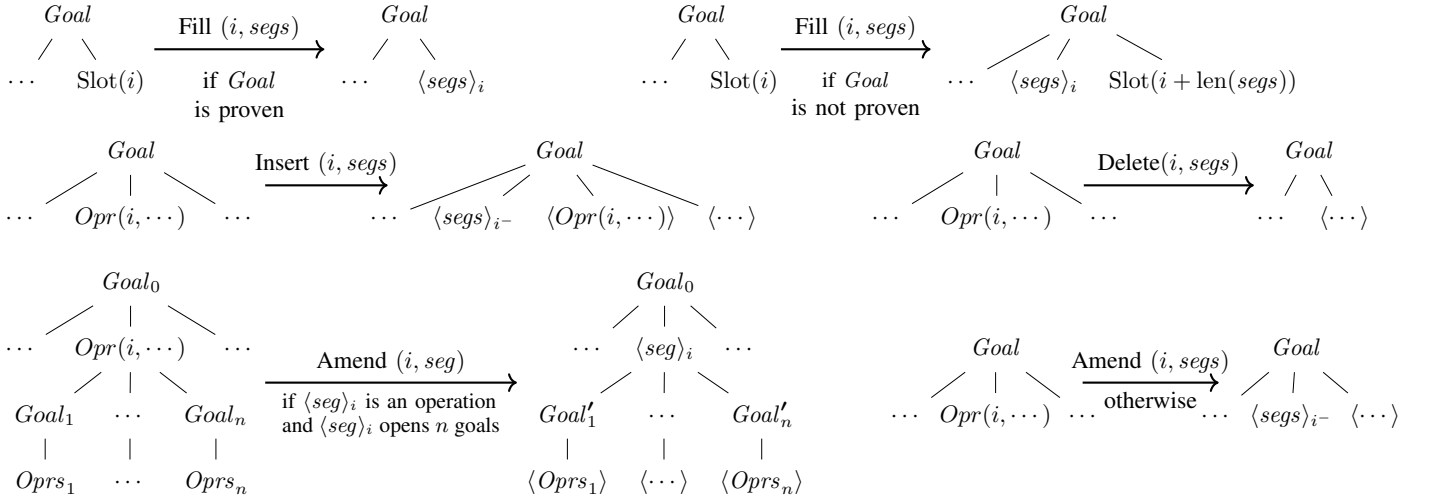


\subsubsection{Tree Editing}\label{sec:edit-model}
The concrete implementation of the \textsc{Edit} tool takes three arguments: the identifier of the target node, an edit mode (fill/insert/amend/delete), and a JSON representation of Minilang ASTs.
For ease of presentation, we use the following abstract syntax for \textsc{Edit} calls:
\begin{align*}
\mathrm{Edit\_Tool} \Coloneqq&~\mathrm{Fill}\,(id, \mathit{segs}) \qquad \mathit{segs} \in \mathrm{Minilang\_Seg}^+\\*
\mathrel{|}&~\mathrm{Insert\_Before}\,(id, \mathit{segs})\\*
\mathrel{|}&~\mathrm{Amend}\,(id, \mathit{segs})\\*
\mathrel{|}&~\mathrm{Delete}\,(id)\\
\mathrm{Minilang\_Seg} \Coloneqq&~\mathrm{Minilang\_Cmd}\ \text{extended with}\ \mathrm{Hole}
\end{align*}
A $\mathrm{Hole}$ is a nullary Minilang operator that marks the point where the current edit stops, leaving the rest of the proof to a subsequent edit; when the edit is applied, each $\mathrm{Hole}$ is realized as a $\Slot$ in the proof-script tree.

The behaviors of the four edit commands is illustrated in \cref{fig:tree-edit}.
$\mathrm{Fill}\,(id, \mathit{segs})$ replaces the $\Slot$ identified by $id$ with the given sequence of operations; should these not yet close the goal that owned the slot, a fresh $\Slot$ is left after them so that the proof can be continued.
$\mathrm{Insert\_Before}\,(id, \mathit{segs})$ splices the given operations into the proof immediately before node $id$, leaving that node and its siblings in place. 
$\mathrm{Amend}\,(id, \mathit{segs})$ usually deletes the subtree at $id$ and replaces it with the new operations; specially, if $\mathit{segs}$ is a single operation that, applied to the same goal, opens exactly as many subgoals as the existing node $id$ did, then only node $id$ itself is replaced in place, and each subgoal retains the proof it already carried.
The latter special semantics of Amend is helpful when one wishes to amend merely a node's operator or arguments, sparing the need to delete its entire subtree and rebuild it from scratch.
Finally, the last command is $\mathrm{Delete}\,(id)$, which removes node $id$ and its entire subtree.

\newsavebox{\treeMinilang}
\begin{lrbox}{\treeMinilang}
\begin{minipage}{5cm}\centering
\begin{tikzpicture}[scale=0.9, transform shape]
  \node (X) at (0,0) {%
    \begin{forest} for tree={align=center, l sep=0, s sep=0.1cm}
      [{$\mathrm{Opr}(i, Op, \mathit{args}, \mathit{msg})$}
        [{$\mathrm{Goal}(i.1, s_1)$} [$\xeval{\mathit{segs}_1}_{i.1.1}$]]
        [$\cdots$ [$\cdots$]]
        [{$\mathrm{Goal}(i.n, s_n)$} [$\xeval{\mathit{segs}_n}_{i.n.1}$]]
      ]
    \end{forest}
  };
\end{tikzpicture}
\end{minipage}
\end{lrbox}

\subsubsection{Proof Evaluation}\label{sec:proof-evaluation}
In \cref{fig:tree-edit}, notation $\xeval{\mathit{node}}$ denotes re-evaluating the $\mathit{node}$ together with its subtree, refreshing the proof state and messages recorded in the $\mathit{msg}$ field of each.
This means that the edited node and all the siblings following it are re-evaluated after every edit, propagating the edit's effect on the proof state through the rest of the proof in real time.
Owing to Minilang's language design, the effect of each proof operation on the proof state is confined to the proof block in which it resides; it therefore suffices to re-evaluate only the siblings following it, leaving all external nodes untouched.

The exact formalization of this re-evaluation mechanism and the definition of $\xeval{\mathit{node}}$ lie outside the main novelty of this paper; we therefore defer them to the released source code in our supplementary material.
We should note, nonetheless, that the mechanism is non-trivial in handling evaluation failures. In our implementation, each node additionally carries an evaluation status, either successful, failed, or pending. When a node's operation fails to evaluate, all the siblings following it are marked pending, and their proof states, along with the prover's feedback recorded in $\mathit{msg}$, become unavailable.

As another related notation in \cref{fig:tree-edit}, the newly added nodes are denoted $\xeval{\mathit{segs}}_i$, with $i$ the identifier at which they are placed.
Function $\xeval{\cdot}_i :: \mathrm{Minilang\_Seg}^+ \rightarrow \mathrm{Node}^+$ elaborates the segments $\mathit{segs}$ into proof-script tree nodes and evaluates the nodes.
Since we do not formalize the evaluation mechanism in this paper, we provide only a partial definition of $\xeval{\mathit{segs}}_i$, which ignores the case of evaluation failure.
Assuming the operation $\mathit{Op}(\mathit{args})$ evaluates successfully on the proof state at $i$ and opens $n$ subgoals with statements $\{s_k\}_{k=1}^{n}$,
\begin{gather*}
\HugeLang{3}{7}
\begin{aligned}
&Op(\mathit{args})\\
&\hspace{3em}\mathit{segs}_1\\
&\texttt{NEXT}\ \cdots\\
&\texttt{NEXT}\ \mathit{segs}_n\\
&\texttt{END}
\end{aligned}
\HugeRang{3}{7}_i
\triangleq \usebox{\treeMinilang}
\\[.3em]
\xeval{Op(\mathit{args})}_i \triangleq \mathrm{Opr}(i, Op, \mathit{args}, \mathit{msg}) \qquad
\xeval{\mathrm{Hole}}_i \triangleq \Slot(i)\\
\xeval{\mathit{seg}_1, \cdots, \mathit{seg}_n}_i \triangleq \xeval{\mathit{seg}_1}_i, \cdots, \xeval{\mathit{seg}_n}_{i+(n-1)}
\end{gather*}
Following Minilang's structural hierarchy, the function $\xeval{\cdot}_i$ elaborates each Minilang operator into an $\Opr$ node, evaluates that operation, records the prover's feedback in its $\mathit{msg}$ field, spawns the resulting subgoals as $\Goal$ nodes, and recurses on the proof of each subgoal.
This recursion continues until it reaches an atomic operation with no subgoals, or a $\mathrm{Hole}$, which is elaborated into a $\Slot$.





\subsection{The Proof Agent Built on top of the Tree-Edit Model}\label{sec:agent}

Given the tree-edit proof model formalized in the previous subsection, building a proof agent on top of it is largely straightforward, amounting to little more than implementing its edit commands as LLM tools.

The Read tool takes an $\mathit{id}$ as its argument, serializes the subtree rooted at node $\mathit{id}$ as text, and returns the text to the LLM. One design choice here is how that subtree is serialized. We simply choose YAML because 1) it is naturally suited to representing trees, 2) it is readable, and 3) it is well represented in LLMs' training corpora.

Another design question concerns the Edit tool's response, which must convey the resulting changes in proof state back to the LLM economically. Naively returning the entire proof tree carries a great deal of redundant information, especially since each edit affects only a subtree. We therefore return a compressed representation of the tree (\cref{fig:compressed_proof_tree}). For each executed operation, it shows only the prover's feedback recorded in the $\mathit{msg}$ field, omitting the operation's $\mathit{args}$, which the LLM itself supplied and which already appear in the tool-calling records in its context. The resulting proof goal is printed only for operations that actually change it, so an operation such as \minilang{Have} (\cref{fig:minilang_example}, line 10) contributes none. Finally, any completed subtree free of evaluation failures is collapsed, since a successfully proven branch rarely needs further attention.

\section{Evaluation}

Having presented the AoA agent and the tree-edit proof model it is built on, we now evaluate whether the design delivers on the two research questions raised in \cref{sec:Intro}.
\begin{enumerate}
\item[\textbf{RQ1.}] How can we reduce the API cost of proof agents, from the side of agent design?
\item[\textbf{RQ2.}] Can we enable general-purpose LLMs to construct formal proofs in a language they have rarely or never encountered, purely through system design and without modifying the underlying models?
\end{enumerate}
We accordingly organize our evaluation around these two questions, first measuring the API cost and token consumption of AoA against a conventional agent, Amazon's Isabelle Agent (RQ1), then testing whether it can prove theorems in Minilang with a model predating the public release of the Minilang proof corpus (RQ2).

\subsection{RQ1: Cost of Proof Agents}\label{sec:eval_RQ1}

To answer RQ1, we compare the cost of AoA against that of a representative conventional proof agent, Amazon's Isabelle Agent. We first measure the overall difference in cost between the two, and then analyze where that difference originates, separating the contribution of our tree-edit interaction design from that of the proof language itself.

\subsubsection{Background}\label{sec:amazon_agent}

Although most existing proof agents are built for Lean, AoA runs on Isabelle/HOL, the prover platform on which Minilang is built. This makes the choice of baseline delicate. A direct comparison with a Lean-based agent would not only compare two agent designs, but also two different prover ecosystems: Lean and Isabelle come with different proof languages, automation facilities, libraries, and interaction mechanisms. Such a comparison would therefore conflate the effect of AoA's agent design with that of the underlying prover. To obtain a cleaner comparison, we instead evaluate AoA against an agent on the same platform, Amazon's Isabelle Agent. This holds Isabelle/HOL and its automation ecosystem fixed, allowing the comparison to focus more directly on the proof language, interaction model, and agent architecture.

Amazon's Isabelle Agent is a strong same-platform baseline for this comparison. Developed by Amazon's Automated Reasoning Group as part of the open-source AutoCorrode verification framework~\cite{AutoCorrode}, it is an LLM agent for interactive proof development in Isabelle/HOL.
The system is professionally engineered: it is integrated with Isabelle’s internals, and provides tools for proof-state inspection, theorem search, and Sledgehammer~\cite{Sledgehammer}, Isabelle's powerful automated proof tool.
It therefore represents a mature instance of the conventional proof-agent design, in which the model edits proof text in Isar~\cite{Isar}, Isabelle's original proof language, and retrieves the surrounding proof context through separate tool calls.
We use it as the baseline against which to evaluate AoA's tree-edit interaction model.

\subsubsection{Benchmarks}

We evaluate on two complementary benchmarks. The first is miniF2F~\cite{minif2f}, a standard benchmark for LLM-based automated theorem proving. Although miniF2F primarily consists of competition-style mathematical problems, such as those from the IMO and AIME, it has become one of the most widely used benchmarks for evaluating proof agents. We therefore include it as a representative benchmark for assessing general theorem-proving capability.

The second benchmark is NTP4VC-Pearl~\cite{NTP4VC}, which we use to evaluate proof agents on verification-condition proving. NTP4VC-Pearl consists of verification conditions generated from verified programs written in Why3, a mature and widely used industrial platform for deductive program verification.
The benchmark covers compact but proof-intensive verification examples, including algorithms, data structures, calculation-heavy programs, engineering-style examples, and verification-competition problems.
We focus on Pearl because it provides a realistic VC structure while remaining small enough for controlled agent evaluation, and because its low ATP pass rate makes it a challenging testbed for measuring whether proof agents can reduce the manual proof burden in program verification.

\subsubsection{Experimental Setup}\label{sec:eval_setup}
We run both AoA and Amazon's agent on the same Isabelle/HOL distribution, Isabelle2025-2, and allocate 16 CPU cores to each run. This configuration is intended to reflect a realistic desktop deployment scenario. For both agents and on both benchmarks, each problem is given a timeout of three hours.

Both agents are allowed to invoke all of Isabelle's proof automation, including Sledgehammer. Since Amazon's agent is optimized for Claude models, we prioritize Claude Opus 4.8 with high thinking effort as the underlying language model for both agents. This removes the language model as a confounding factor and lets the comparison focus on the agent design, proof language, and interaction model. We note that Amazon's agent provides a memory mechanism, whereas the current implementation of AoA does not.

The two agents differ in how they access the LLM API. AoA uses Claude Code as its API provider and is run under a subscription-based plan, for which we purchased \$600 of membership. Because Claude Code's subscription pricing substantially discounts API usage, we do not impose an explicit API-cost cap on AoA.
In contrast, Amazon's agent cannot be run through Claude Code in its current implementation and must instead call the Claude API directly. Since this leads to substantially higher API cost, we impose a total budget of \$3000 for Amazon's agent alone.
Note that all costs reported in this paper are computed from token usage at the standard per-token prices, independent of how each agent is billed.
For Claude Opus, we normalize all input-cache writes to the 5-minute tier, so that price differences between cache-write tiers do not confound the comparison.

Even under these limits, evaluating Amazon's agent on the full benchmarks is beyond our available budget. We therefore evaluate it on randomly sampled subsets: 73 miniF2F problems, covering approximately 30\% of the benchmark, and 50 NTP4VC-Pearl problems, covering approximately 18\% of the Pearl benchmark cases. The sampling ratio is lower for NTP4VC-Pearl because this benchmark contains many difficult verification conditions, which are expensive for both AoA and Amazon's agent to attempt.

\begin{table}[t]
\centering
\caption{Evaluation of AoA and Amazon's agent.}
\label{tab:pass-rate}
\setlength{\tabcolsep}{5pt}
\begin{tabular}{c c c c c c}
\toprule
  \bf agent & \bf model & \bf bench & \bf token & \bf cost & \bf pass \\\midrule
  AoA & Opus 4.8 & miniF2F (all) & 0.68M & \$1.4 & 99.6\% \\
  Amazon's & Opus 4.8 & miniF2F (73) & 3.27M & \$17.2 & 89.0\% \\
  AoA & Opus 4.5 & miniF2F (73) & 1.44M & \$1.5 & 98.6\% \\
  AoA & GPT 5.5  & miniF2F (all) & 0.19M & \$0.4 & 99.2\% \\
  AoA & GPT 6 sol & miniF2F (all) & 0.27M & \$0.17 & \bf 100\%
\\\midrule
  AoA & Opus 4.8 & NTP4VC-P (31) & 9.07M & \$13.9 & 93.6\% \\
  Amazon's & Opus 4.8 & NTP4VC-P (31) & 8.28M & \$32.0 & 48.4\% \\\midrule
  AoA & GPT 5.5 & NTP4VC-P (all) & 2.42M & \$3.9 & 89.2\% \\
  AoA & GPT 5.5 & NTP4VC-P (50) & 2.56M & \$4.2 & 92.0\% \\
  Amazon's & GPT 5.5 & NTP4VC-P (50) & 4.03M & \$12.7 & 38.0\% \\
AoA & GPT 5.5 & {\footnotesize NTP4VC-RealC} (all) & 0.80M & \$1.5 & 97.7\%
\\\bottomrule
\end{tabular}
\footnotesize\flushleft
\vspace{-.6em}
\item All the metrics are averaged per case, considering only \textit{passed} cases.
\end{table}

\begin{table*}[t]
\centering
\caption{Comparing Detailed Metrics of AoA and Amazon's agent on the common success sets.}
\setlength{\tabcolsep}{4.5pt}
\label{tab:results}
\begin{tabular}{c c c c c c c c c c c}
\toprule
  \bf agent & \bf model & \bf bench. & \!\!\!\!\bf note\!\!\!\! & \bf total tok. & \bf input tok. & \bf cached & \bf \!\!output tok.\!\! & \bf cost & \bf \# tool & \bf time (min) \\\midrule
  AoA & \multirow{3}{*}{Opus 4.8} & \multirow{3}{13.5mm}{\centering  miniF2F\\(common success)} & - &
  0.48M (1x) & 0.47M (1x) & 0.41M (1x) & 18k (1x) & \$1.0 (1x) & 17.7 (1x) & 6.2 (1x)
\\
  Amazon's &  & & O &
  3.34M (6.9x) & 3.30M (7.1x) & 0 & 40k (2.3x) & \$17.5 (17.9x) & 157.3 (8.9x) & 12.2 (2.0x)
\\
  Amazon's &  & & N &
  - & - & 2.87M (6.9x) & - & \$4.6 (4.7x) & - & -
\\\midrule
  AoA & \multirow{3}{*}{Opus 4.8} & \multirow{3}{*}{\parbox{1.35cm}{\centering NTP4VC-P\\(common success)}} & - &
  2.44M (1x) & 2.37M (1x) & 2.24M (1x) & 65.7k (1x) & \$3.6 (1x) & 37.3 (1x) & 18.1 (1x)
\\
  Amazon's & & & C &
  8.28M (3.4x) & 8.21M (3.5x) & 2.41M (1.1x) & 70.5k (1.1x) & \$32.0 (9.0x) & 239.4 (6.4x) & 24.7 (1.4x)
\\
  Amazon's & & & N &
  - & - & 7.72M (3.5x) & - & \$8.1 (2.3x) & - & -
\\\midrule
  AoA & \multirow{3}{*}{GPT 5.5} & \multirow{3}{*}{\parbox{1.35cm}{\centering NTP4VC-P\\(common success)}} & - &
  1.41M (1x) & 1.39M (1x) & 1.17M (1x) & 18k (1x) & \$2.2 (1x) & 50.3 (1x) & 16.5 (1x)
\\
  Amazon's & & & C &
  4.03M (2.9x) & 3.98M (2.9x) & 1.92M (1.6x) & 49k (2.7x) & \$12.7 (5.7x) & 194.0 (3.9x) & 32.1 (1.9x)
\\
  Amazon's & & & N &
  - & - & 3.30M (2.8x) & - & \$6.6 (2.9x) & - & -
\\\bottomrule
\end{tabular}
\footnotesize\flushleft
\vspace{-.6em}
\item Note: O denotes the original result of Amazon’s agent; N denotes the normalized result under the same input-cache rate as AoA; C denotes our modified implementation with inserted cache-control directives. All the metrics are averaged per case.
``-'' symbol in N rows means the digit is unchanged.
\end{table*}

\subsubsection{Results on miniF2F}\label{sec:miniF2F_results}

The results are reported in \cref{tab:pass-rate} and \cref{tab:results}.
On the selected 73 miniF2F problems, AoA solves all cases, while Amazon’s agent solves 65 cases (89.0\%). \cref{tab:results} also reports, for each agent, the average per-case total token consumption, input tokens, cached input tokens, output tokens, API cost, and elapsed time.

One caveat in comparing API cost is input-cache reuse. In the original runs, Amazon’s agent records zero cache, because its implementation enables the caching feature only on Amazon's Bedrock platform rather than the Anthropic API provider we use.
Because cached input is substantially cheaper, the original API costs conflate agent efficiency with cache reuse.
To remove this confounding factor, we also report a normalized version of Amazon’s results (denoted ``N'' in \cref{tab:results}), where we assume that Amazon’s agent achieves the same input-cache rate as AoA, and recompute its cached input tokens and resulting API cost.

For a more direct efficiency comparison, we further restrict attention to the common success set, namely the miniF2F problems solved by both agents, since statistics from failed runs do not directly measure the cost of completing a proof.
The results show a substantial efficiency advantage for AoA across all the economic metrics. On the common success set, Amazon’s agent consumes $6.9{\times}$ as many total tokens as AoA. Even after normalizing away the cache-reuse difference, Amazon’s API cost remains $4.7{\times}$ that of AoA. Moreover, AoA's advantage is not limited to monetary cost: AoA also completes the common successful cases in roughly half the elapsed time of Amazon’s agent, taking 371 seconds per case compared with Amazon’s 730 seconds.

\begin{figure}[t]
    \centering
    \includegraphics[width=1\linewidth]{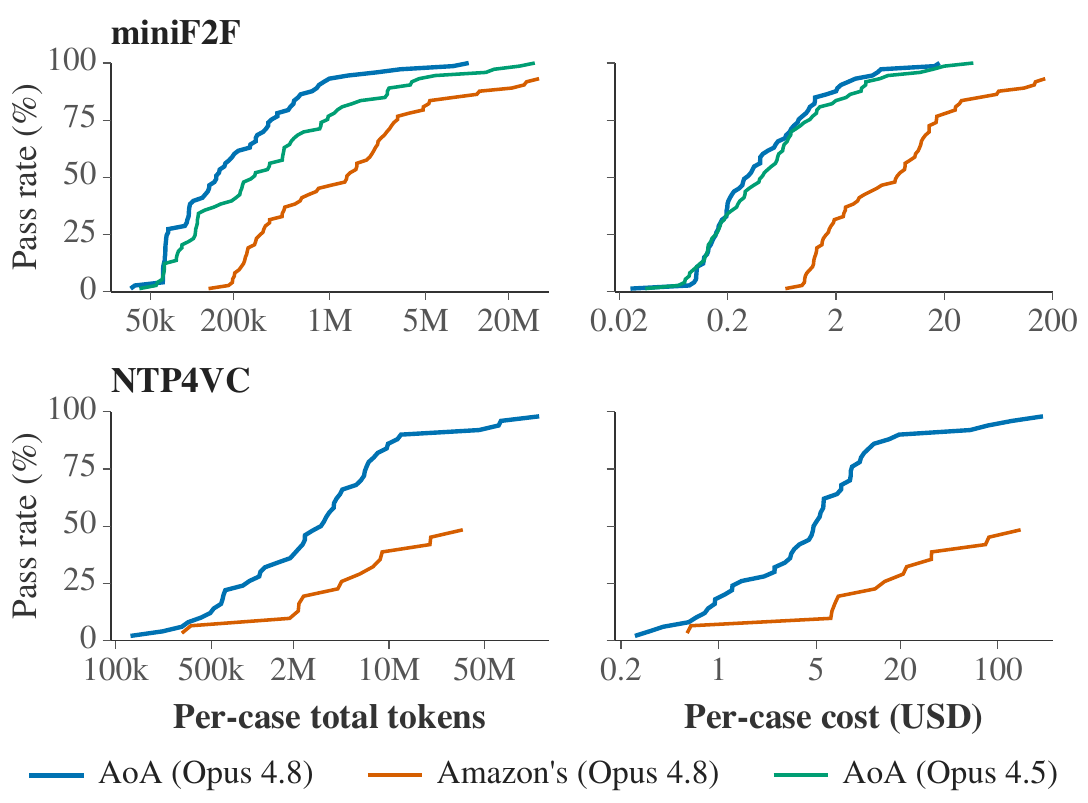}
    \caption{Pass rate under increasing per-case budgets.}
    \label{fig:pass_vs_token}
\end{figure}
\subsubsection{Results on NTP4VC}\label{sec:NTP4VC_results}

The results on the 50 NTP4VC-Pearl problems are also reported in \cref{tab:pass-rate} and \cref{tab:results}.
This benchmark is substantially harder than miniF2F and is therefore much more expensive for Amazon’s agent to evaluate.
To make the evaluation more cost-efficient, we modified Amazon’s agent to set \verb|cache_control| in its API requests, thereby enabling the API provider’s automatic input-cache mechanism~\cite{anthropic-pc}. Even with this modification, however, the cost of evaluating Amazon’s agent had already exceeded USD 1,500 after completing only 31 cases, forcing us to stop the Claude-based experiment early.
Instead, we completed the full 50-case evaluation using GPT 5.5, for which we had access to substantially larger quota. This full evaluation, including the runs of AoA, consumed about 1B tokens in total.

The results show that AoA substantially outperforms Amazon’s agent in pass rate, while solving all cases solved by Amazon’s agent.
As in the miniF2F analysis, to ensure a fair comparison, we again compare the two agents on their common success set (\cref{tab:results}).
The results exhibit the same overall pattern as on miniF2F: AoA achieves multi-fold improvements across all metrics, while also finishing faster.


\subsubsection{Analysis}\label{sec:eval_analysis}

We further analyze the source of AoA’s advantage. A natural question is whether this advantage depends on the difficulty of the proof goals. To examine this, we plot the cumulative pass rate as a function of the per-case budget in \cref{fig:pass_vs_token}.
Each point in the plot reports the fraction of problems solved within a given token budget and a tool-call number limit. Read inversely, the plot shows how the required budget grows as we move to increasingly difficult proof goals.
On miniF2F, where Amazon’s agent eventually performs well, the budget gap between the two agents remains relatively uniform as we move toward higher pass-rate levels. On NTP4VC, however, the gap becomes much larger at harder proof goals. These results suggest that harder proof goals can amplify AoA’s advantage.

A second question is which part of AoA's design this advantage comes from.
AoA differs from Amazon's agent in two respects at once: it uses Minilang instead of Isar, and it replaces source-text editing with the tree-edit model.
The improvements shown above thus combine the effects of both.
Separating the two, however, is difficult.
On the one hand, building a conventional text-editing agent for Minilang and comparing it with an Isar-based agent would be unfair to Minilang, because it is a new language and general-purpose LLMs do not have the same prior familiarity with it as they do with Isar.
On the other hand, removing Minilang and applying the tree-edit model directly to Isar is also difficult.
Unlike Minilang, whose proof model is kept uniformly declarative, Isar interleaves declarative proof steps, tactical goal refinement, and fact chaining; fully encoding this multi-mode proof model, together with Isar’s rich and diverse syntax, into a JSON schema would make the interface large and unwieldy, likely undermining the LLM efficiency that the tree-edit model is meant to improve.

\begin{table}[t]
\centering
\caption{AoA vs. Minilang's text-editing agent (TextEd)}
\label{tab:ablation}
\setlength{\tabcolsep}{2.7pt}
\begin{tabular}{c c c c c c}
\toprule
  \bf agent & \bf bench. & \bf total tok. & \bf input tok. & \bf output tok. & \bf cost \\\midrule
  AoA & \multirow{2}{13mm}{\centering miniF2F\\(common)} & 0.53M (1x) & 0.50M (1x) & 22k (1x) & \$1.1 (1x) \\
  TextEd & & 1.19M\,(2.3x) & 1.17M\,(2.3x) & 18k\,(0.8x) & \$1.8\,(1.6x) \\
\midrule
  AoA & \multirow{2}{13mm}{\centering NTP4VC\\(common)} & 7.37M (1x) & 7.25M (1x) & 117k (1x) & \$10 (1x) \\
  TextEd & & 28.5M\,(3.9x) & 28.3M\,(3.9x) & 185k\,(1.6x) & \$28 (2.7x)
\\\bottomrule
\end{tabular}
\footnotesize\flushleft
\vspace{-.6em}
\item 
Metrics are per-case averages over the common success set; the text-editing agent's input-cache rate is normalized to AoA's.
Model: Claude Opus 4.8.


\end{table}

Although these difficulties preclude a clean ablation, we still build a conventional-style text-editing proof agent for Minilang as a reference point. It teaches the LLM Minilang through in-context learning, by providing the grammar and examples in its system prompt.
Comparing it with AoA holds the proof language (Minilang) fixed, and thus separates the two factors to some extent.
We evaluate it with Claude Opus 4.8 on the same 73 miniF2F cases and 31 NTP4VC-Pearl cases used earlier in this section; the results are presented in \cref{tab:ablation}.
First, AoA achieves a higher pass rate: on the miniF2F subset, AoA passes 100\% of the cases and the text-editing agent 98.6\%; on the NTP4VC-Pearl subset, 93.5\% and 80.6\%, respectively.
Moreover, on the common success sets, AoA consumes substantially fewer tokens and incurs a lower API cost than the text-editing agent.
These results indicate that the tree-edit model indeed helps the LLM use Minilang more effectively, in both pass rate and efficiency.

A more interesting observation is that, on the miniF2F subset, the text-editing agent actually produces fewer output tokens than AoA.
This is expected: on easy problems, proofs need little revision, so the output consists mostly of the proof itself, and the JSON representation of a proof AST is indeed longer than the original proof script and thus takes more tokens.
On harder problems, however, such as those in the NTP4VC-Pearl subset, LLMs must revise the proof repeatedly, and the two agents diverge.
On miniF2F, the text-editing agent makes 5.7 proof edits per case on average, about the same as AoA's 5.6; on NTP4VC-Pearl, it makes 40.7, 1.44$\times$ AoA's 28.3.
The rest of the gap in output tokens comes mainly from additional thinking.
On these hard problems, the advantage of the tree-edit model over the whole interactive proof-construction process comes into play, and AoA regains the lead in output-token consumption.


More generally, \cref{tab:results} and \cref{tab:ablation} both show that AoA's reduction in input tokens is consistently larger than that in output tokens.
This matches the intuition behind our design: conventional agents spend many tokens repeatedly navigating source files and querying proof states, whereas AoA exposes the same information more effectively through the structured proof tree.
To examine this mechanism more directly, we measure, over the entire LLM session, the input-token footprint of the tool outputs used for source-code reading and proof-state access.
For Amazon’s agent, we count the outputs of tools that read proof source and inspect proof states (\verb|get_goal_state|, \verb|get_errors|, \verb|read_theory|, ... the full list is in our supplementary material). For AoA, we count the outputs of \textsc{READ}, which returns the complete proof tree or a requested subtree, and \textsc{EDIT}, whose response contains the incremental proof-state quick view.
The results are consistent with this explanation.
On the common success set of NTP4VC with GPT 5.5, Amazon’s relevant tool outputs consume 2.6M input tokens per problem on average (63.2\% of the total input), whereas AoA’s consume only 0.47M (34.0\%).
This means that AoA not only reduces the absolute input-token cost of communicating proof context to the model, but also leaves a larger fraction of the input budget for the model’s own reasoning history.
This reduction accounts for part of AoA’s overall advantage in \cref{tab:results}.


Putting together the results in \cref{sec:miniF2F_results}, \cref{sec:NTP4VC_results} and the analysis in this subsection, we now answer RQ1.

\textbf{Answer to RQ1.} 
Redesigning the proof language and exposing its AST through a JSON representation within a tree-edit proof model provides an effective way to reduce proof-agent cost. AoA serves as a concrete instance of this approach.




\subsection{RQ2: Proving in Minilang Without Corpus-Scale Exposure}\label{sec:eval_unseen}

The experiments in the previous section already provide evidence that AoA can operate effectively over Minilang with Claude Opus 4.8 and GPT 5.5. However, they do not fully rule out a possible data-contamination explanation. Minilang itself comes with a corpus of about 340K proofs, released publicly on Hugging Face~\cite{minilang_afp_v1}. It is therefore difficult to exclude the possibility that those LLMs have absorbed part of this corpus during training, and have thereby acquired some familiarity with Minilang.

To address this concern, we conduct an additional experiment on miniF2F using Claude Opus 4.5, which was released on Nov. 24, 2025, before the public release of the Minilang corpus on Dec. 22, 2025~\cite{minilang_afp_v1}. This setting therefore rules out training on the publicly released Minilang corpus.

We evaluate Claude Opus 4.5 on the same selected miniF2F cases and under the same setting as \cref{sec:eval_RQ1}. 
The results are reported in \cref{tab:pass-rate} and \cref{fig:pass_vs_token}. AoA still achieves a pass rate of 98.6\%, even though the underlying model predates the public release of the Minilang corpus.
If we extend the time limit to 4.5h, the pass rate reaches 100\%.
Moreover, its resource statistics, including total token consumption and API cost, are not excessively worse than those obtained with Claude Opus 4.8.
Considering that Opus 4.5 and Opus 4.8 are separated by three model versions and roughly half a year of development, the observed differences in these statistics can be reasonably attributed to the capability gap between the underlying language models.

These results provide direct evidence for an affirmative answer to RQ2 in the case of Minilang. 

\textbf{Answer to RQ2.}
At least for Minilang, the answer is affirmative. By extracting the semantics of the proof language and the prover interaction, and re-expressing them in forms more natural for LLMs to manipulate, we can build an effective proof agent even with an underlying LLM that was not trained on a corpus-scale collection of Minilang proofs.

This conclusion should be read with the scope of Minilang in mind. Minilang is designed so that its statements and proof steps stay close to natural-language mathematical reasoning.
The experiment therefore does not by itself show that the same approach applies to arbitrary formal languages, especially those farther from reasoning patterns familiar to LLMs.

Nevertheless, the result provides a useful direction for the design of proof agents for new languages. It suggests that, when developing proof languages and formal languages more broadly, one can make them more amenable to LLM agents by exposing their semantic structure through explicit, structured, and LLM-friendly interfaces, rather than relying solely on model retraining or prior exposure to the concrete syntax.




\subsection{Threats to Validity}

Several threats may affect the interpretation of our results. First, our benchmarks may not fully represent all theorem-proving and software-verification workloads. MiniF2F mainly consists of mathematical problems, while NTP4VC-Pearl is closer to verification-condition proving but is derived from Why3 as a single verification framework. AoA may behave differently on VCs generated by other verifiers, with different encoding styles, background theories, or proof-obligation structures.
Second, the high cost of running the baseline limits its evaluation to sampled subsets, which may affect the statistical stability of the aggregate results.
We mitigate this threat by using random sampling, reporting pass rates separately from resource metrics, and comparing resource usage on the common success sets.
Third, our comparison uses Amazon's agent as the same-platform representative of conventional proof agents. This avoids confounding Isabelle/HOL with other prover platforms, but it also means that the results may not directly transfer to other agents.

\section{Related Work}

This section situates AoA with respect to four lines of related work: conventional proof agents, tree-editing techniques for code generation, tree-editing interfaces for interactive theorem proving, and fine-tuning-based theorem-proving models.

\subsection{Conventional Proof Agents}

The conventional proof-agent paradigm has already been characterized in \cref{sec:existing_agent}, so we only list representative systems here. Representative works include checker-in-the-loop proof-script generators \cite{thakur2024an,CoqPilot}, agents that add iterative refinement, memory, library search, or coding-agent infrastructure \cite{requena2026minimal,Numina-Lean-Agent}, and more recent systems that incorporate decomposition and planning \cite{Prover-Agent,Archonn,LEAP}.
In Isabelle/HOL, Amazon's Isabelle Agent \cite{AutoCorrode} provides the closest same-platform representative of this conventional paradigm, as discussed in \cref{sec:existing_agent}.

Across provers, the previous state of the art among agentic provers on miniF2F is FormalInception~\cite{formalinception2026minif2f}, which reports solving all 244 problems, as the union of several runs with different agent settings, at \$0.35 per case under Gemini's half-price API rates.
AoA with GPT~6~sol solves all of them in a single run at \$0.17 per case at standard rates.
On NTP4VC-Pearl, whose pass rates are directly comparable across Lean, Rocq, and Isabelle, AoA's 89.2\% is, to our knowledge, the highest reported; no prior agent has reported its cost.

\subsection{AST-Based Program Generation and Agents} 

A related line of work in program generation uses the abstract syntax of the target language as the model’s output interface, rather than asking the model to directly emit a sequence of concrete-syntax tokens. Representative systems such as Abstract Syntax Networks~\cite{rabinovich-etal-2017-abstract}, Tree-to-Tree networks~\cite{3327144.3327180}, TRANX~\cite{TRANX}, and TreeGen~\cite{TreeGen} generate programs through AST nodes, grammar actions, or grammar-aware tree structures, thereby turning program construction into a sequence of language-level structural decisions rather than unconstrained surface-token prediction. Another related line of work aligns compiler feedback with structured program representations. For example, program-repair systems such as Graph2Diff~\cite{Graph2Diff} and DrRepair~\cite{DrRepair} connect compiler diagnostics or error messages with code structures, showing that feedback from language tools can be incorporated into the representation on which a model reasons. Finally, recent coding agents~\cite{kimi-k2,glm-5} demonstrate the usefulness of closing the loop with tools, compiler messages, tests, and repository navigation. Together, these works show that program-generation systems can benefit both from exposing abstract syntax to the model and from making external tool feedback available during generation or repair.
AoA is inspired by these ideas, but applies them to interactive theorem proving.

\subsection{Tree-Structured Proofs in Interactive Theorem Proving}

Tree-structured proofs have a long history in logic and interactive theorem proving. In proof theory, natural deduction and sequent calculus organize derivations as trees of inference-rule applications, whose branches are subderivations.

This perspective has also influenced interactive proof environments. Proof by Pointing~\cite{ProofByPointing} uses locations in a goal sequent to guide the construction of a proof tree. ProofViz~\cite{ProofViz} visualizes proof trees for Cur with node information and partial proof terms, supporting interactive proof exploration. HenBlocks~\cite{BoeyAdams2022HenBlocks} further explores structured proof construction through a Blockly-based editor for Coq. These systems are interfaces for human users rather than LLM agents.
AoA builds on this broader tradition of tree-structured proof presentation and construction, but exposes the tree as the working object of an LLM agent, with proof states and prover messages stored at the nodes that the agent reads and edits.

\subsection{Finetuning-based Theorem Proving Models}
As an LLM-powered approach to theorem proving, AoA is also related to the growing line of fine-tuning-based theorem-proving models.
Recent theorem-proving models are primarily fine-tuned for Lean, including Pythagoras-Prover~\cite{Pythagoras-Prover}, Goedel-Prover-V2~\cite{Goedel-Prover-V2}, DeepSeek-Prover-V2~\cite{Deepseek-Prover-V2}, and others~\cite{kimina-prover-review, Oprover, Longcat-Prover, STP}. These models are typically trained with specialised Lean-oriented data synthesis pipelines, followed by reinforcement learning. Recent work has also used Lean-verified instances to improve informal reasoning~\cite{leang-etal-2025-theorem}. However, these models mainly excel on certain benchmarks such as miniF2F~\cite{minif2f}, while struggling to generalise to broader complex benchmarks~\cite{putnam, FATE, proofnet, NTP4VC}.
\section{Conclusion and an Open Problem}

This paper presents AoA, a proof agent built on a tree-edit interface over Minilang’s abstract syntax tree. Through AoA, we show that proof agents need not necessarily rely on corpus-scale prior familiarity with a proof language’s concrete syntax: by exposing the language’s semantic structure in an LLM-friendly form, the efficiency benefits of a redesigned proof language can be extended to general-purpose, non-finetuned LLM agents, substantially lowering the API cost.

This work focuses on a proof language. Given the close analogy between proof languages and programming languages, an open question is whether the same agent-over-AST methodology can also benefit coding agents, especially for newly designed experimental languages whose concrete syntax is not yet familiar to existing LLMs.
We leave this for future work.


\section*{Acknowledgement}

Claude Fable 5 series are used to refine the language and improve the readability of this article. 
Conrad Watt and Qiyuan Xu are supported by the Ministry of Education, Singapore, under its Academic Research Fund Tier 1 (NTU Tier 1, RG18/25). Conrad Watt and Qiyuan Xu are additionally supported by an NTU Nanyang Assistant Professorship Start-Up Grant.
Wenda Li and Joshua Ong Jun Leang acknowledge the support of Renaissance Philanthropy through the AI for Math Grant.

\newpage

\bibliographystyle{IEEEtran}
\bibliography{bib}

\end{document}